\documentclass{article}
\usepackage{graphicx}
\usepackage{latexsym}
\usepackage[cp1250]{inputenc}
\newcommand{\M}{{\cal M}}

\newcommand{\h}{\hspace{.5 em}}

\newcommand{\bdm}{\begin{displaymath}}
\newcommand{\edm}{\end{displaymath}}
\newcommand{\bi}{\begin{itemize}}
\newcommand{\ei}{\end{itemize}}
\newcommand{\benum}{\begin{enumerate}}
\newcommand{\eenum}{\end{enumerate}}
\newcommand{\be}{\begin{equation}}
\newcommand{\ee}{\end{equation}}
\newcommand{\bea}{\begin{eqnarray}}
\newcommand{\eea}{\end{eqnarray}}
\newcommand{\btabular}{\begin{tabular}}
\newcommand{\etabular}{\end{tabular}}
\newcommand{\beas}{\begin{eqnarray*}}
\newcommand{\eeas}{\end{eqnarray*}}

\newtheorem{defin}{Definition}[section]
\newcommand{\bdefin}{\begin{defin}}
\newcommand{\edefin}{\end{defin}}
\newtheorem{theorem}{Theorem}[section]
\newcommand{\bthrm}{\begin{theorem}}
\newcommand{\ethrm}{\end{theorem}}
\newtheorem{proposition}[defin]{Proposition}
\newcommand{\bproposition}{\begin{proposition}}
\newcommand{\eproposition}{\end{proposition}}
\newtheorem{corollary}[defin]{Corollary}
\newcommand{\bcorollary}{\begin{corollary}}
\newcommand{\ecorollary}{\end{corollary}}
\newtheorem{rem}{Remark}
\newcommand{\brem}{\begin{rem}}
\newcommand{\erem}{\end{rem}}
\newtheorem{Lem}[defin]{Lemma}
\newcommand{\bL}{\begin{Lem}}
\newcommand{\eL}{\end{Lem}}
\newtheorem{exam}{Example}
\newcommand{\bexam}{\begin{exam}}
\newcommand{\eexam}{\end{exam}}
\begin{document}
%
%
\noindent\large $\ast$-SDYM fields and heavenly spaces:  
II. Reductions of the $\ast$-SDYM system. \normalsize\\

\noindent Sebastian Forma\'{n}ski$^{1}$ and Maciej Przanowski$^{1,2}$\\
\mbox{ }\\
\small
$\mbox{ }^{1}$Institute of Physics, Technical University of Ł\'{o}dz,
W\'{o}lcza\'{n}ska 219. 93-005 Łódź, Poland\\
$\mbox{ }^{2}$Departamento de Fisica, Centro de Investigaciones y de Estudios Avanzados del IPN, 
Apartado Postal 14-740, 07000 Mexico DF, Mexico\\

\vspace {1 ex}
\noindent E-mail: sforman@p.lodz.pl and przan@fis.cinvestav.mx
\normalsize
\paragraph{Abstract.} Reductions of self-dual Yang-Mills (SDYM) system for $\ast$-bracket Lie 
algebra to the Husain-Park (HP) heavenly equation and to $sl(N,\mbox{\boldmath{$C$}})$ SDYM 
equation are given. An example of a sequence of $su(N)$ chiral fields ($N\geq 2$) tending 
for $N\rightarrow\infty$ to a curved heavenly space is found.

\vspace {1 ex}

\section*{Introduction} 
This paper is a second part of our previous work (Forma\'{n}ski and Przanowski 2005) 
where the integrability of self-dual Yang-Mills (SDYM) system in the formal 
$\ast$-algebra bundle over heavenly space has been proved. In what follows we refer to the 
first paper as I.
In particular it has been shown in I, that the considered SDYM equations can be reduced to one 
equation called \it master equation \rm (ME)
\be
\label{ME in H}
g^{\tilde{\beta}\alpha}\partial_{\tilde{\beta}}\partial_{\alpha}\Theta+
\frac{1}{2G}\epsilon^{\alpha\beta}\{\partial_{\alpha}\Theta\, ,\, \partial_{\beta}\Theta\}=0
\ee
Notation in the present paper is the same as in the first one. So $g_{\alpha\tilde{\beta}}$ 
are components of the K\H{a}hler metric on the base manifold $\M$ i.e.
$g_{\alpha\tilde{\beta}}=\partial_{\alpha}\partial_{\tilde{\beta}}{\cal K}$, where ${\cal K}$ is 
a K\H{a}hler potential. As $\M$ is a four dimensional heavenly space 
$\det(g_{\alpha\tilde{\beta}})=G(w,z)\tilde{G}(\tilde{w},\tilde{z})$, with 
$\{w,\tilde{w},z,\tilde{z}\}$ being local coordinates on $\M$. The function $\Theta$ takes 
values in  the formal $\ast$-algebra $({\cal A},\ast)$ 
of formal power series over a symplectic manifold $(\Sigma^{2n},\mbox{\boldmath{$\omega$}})$
\be
\label{Theta}
\Theta=\sum_{m=0}^{\infty}\sum^{\infty}_{k=-m}t^{m}\hbar^{k}
\Theta_{m,k}(w,z,\tilde{w},\tilde{z},x^{1},...,x^{2n})
\ee
where $\{x^{1},...,x^{2n}\}$ are local coordinates on $\Sigma^{2n}$ and  
$\Theta_{m,k}(w,z,\tilde{w},\tilde{z},x^{1},...,x^{2n})$ are analytic functions on 
$\M\times\Sigma^{2n}$. The $({\cal A},\ast)$ 
algebra has been discussed in detail in I.

\vspace{.5 ex}

\noindent The evidence of integrability of ME has been achieved by:

\noindent $\bullet$ construction of infinite hierarchy of conservation laws,

\noindent $\bullet$ rewritting the equation (\ref{ME in H}) as an integrability condition for a Lax pair,

\noindent $\bullet$ twistor construction i.e. one-to-one correspondence between solution of (\ref{ME in H})
and vector bundles over twistor space for $\M$.

\noindent $\bullet$ the proof of existence of a solution for a respective Riemann-Hilbert problem.

In the present work we study the role which the equation (\ref{ME in H}) plays among other integrable systems.
As it was discussed previously (cf also Mason and Woodhouse 1996) the SDYM system for finite diemensional
Lie algebras is not general enough to include all lower diemensional integrable systems 
(see also (Ward 1992)). In order
to generalize the SDYM system to obtain a universal one according to the famous \it Ward's conjecture \rm
(Ward 1985) it seems natural to extend the algebra of this system to the one
containing all $sl(N,\mbox{\boldmath{$C$}})$ algebras as well as infinite diemensional algebra of 
Hamiltonian vector fields over $(\Sigma^{2n},\mbox{\boldmath{$\omega$}})$. In this paper we show that 
the algebra chosen in 
(Forma\'{n}ski and Przanowski 2005) meets this requirements. We find a reduction of ME
to the  $sl(N,\mbox{\boldmath{$C$}})$ SDYM equation for $N\geq 2$ as well as to a heavenly 
equation. This also opens the possibiliy to find  a sequence of $su(N)$ chiral fields 
tending to a heavenly space as $N\rightarrow\infty$. The last problem was first posed 
by Ward (1990b).

The paper is organized as follows. Section \ref{Reduction of ME to heavenly equation} is 
devoted to the reduction of  ME to the  Husain-Park (HP) heavenly equation. We concentrate mainly 
on solutions of ME which are analytic functions of deformation parameters $\hbar$ and $t$. 
The example of such a solution is given. The heavenly space obtained is analysed in terms 
of the  null tetrad formalism.

In Section \ref{Reduction of ME to $su(N)$ SDYM system} we consider reduction of ME to the 
$sl(N,\mbox{\boldmath{$C$}})$ SDYM equation. We describe a basis of $sl(N,\mbox{\boldmath{$C$}})$ 
introduced by (Farlie et al 1990). Theorem \ref{o homomorfizmie rozwiazan}, which is a main
result of this section proves that any analytic solution of the $sl(N,\mbox{\boldmath{$C$}})$ 
SDYM equation can be obtained by a reduction of some solution to ME.

Section \ref{From chiral fields to heavenly space} deals with the relations  between 
$sl(N,\mbox{\boldmath{$C$}})$ chiral equation and  ME. We propose a precise meaning  
of limitting process  $N\rightarrow\infty$ for 
$sl(N,\mbox{\boldmath{$C$}})$ chiral fields. An example of $su(N)$ chiral field sequence tending 
to a heavenly space is given.

Concluding remarks close the paper.
%
%
%
\renewcommand{\theequation}{\arabic{section}.\arabic{equation}}
\setcounter{equation}{0}
\section{Reduction of ME to the heavenly equation}
\label{Reduction of ME to heavenly equation}
%
The form of ME (\ref{ME in H}) is quite general. This is due to the fact that 
it is given over an arbitrary heavenly space $\M$ and the algebra in which the 
function $\Theta$ takes values is an arbitrary formal $\ast$-algebra $({\cal A},\ast)$ over
any symplectic manifold $(\Sigma^{2n},\mbox{\boldmath{$\omega$}})$.

In this section we restrict ourselves  to the base manifold $\M=\mbox{\boldmath{$C$}}^{4}$
(or $\mbox{\boldmath{$R$}}^{4}$ of the signature $(++ --)$)
with K\H{a}hler potential ${\cal K}=w\tilde{w}+z\tilde{z}$. Also the symplectic manifold 
is assumed to be two dimensional. Thus ME (\ref{ME in H}) takes the form 
\be
\label{ME in c4}
\partial_{w}\partial_{\tilde{w}}\Theta+\partial_{z}\partial_{\tilde{z}}\Theta+
\{\partial_{w}\Theta\, ,\,\partial_{z}\Theta\}=0.
\ee
We seek for the solutions of the equation (\ref{ME in c4}) which are analytic in all
coordinates and also in both parameters $t$ and $\hbar$. It means that the series 
(\ref{Theta}) is convergent. Then we can interchange the sums i.e. rewrite the 
function $\Theta$ in the form
\be
\label{analytic Theta}
\Theta=\sum^{\infty}_{k=0}\hbar^{k}
\Theta_{k}(t,w,z,\tilde{w},\tilde{z},x^{1},x^{2})
\ee 
where each $\Theta_{k}(t,w,z,\tilde{w},\tilde{z},x^{1},x^{2})$ is analytic in 
all varaibles \mbox{$\{t,w,z,\tilde{w},\tilde{z},x^{1},x^{2}\}$.}

Note, that this case explains why the parameter $t$ has been called in I the \it 
 convergence parameter\rm . Observe also that in many such cases the parameter $t$
can be absorbed by a suitable redefinition of coordinates $\{w,z,\tilde{w},\tilde{z}\}$ on $\M$.

As is known the $\ast$-product has a form of formal power series
\be
\label{ast}
\forall f,g\in C^{\omega}(\Sigma^{2n})\h\h\h f\ast g =\sum_{k=0}^{\infty}\hbar^{k}\Delta_{k}(f,g)
\ee
The analyticity of $\Theta(t,\hbar;w,z,\tilde{w},\tilde{z},x^{1},x^{2})$ in $t$ and $\hbar$ imposes 
strong requirements on the $\ast$-product itself. This product should be also analytic function
of the deformation parameter. Thus the series (\ref{ast}) should be convergent at least for some 
neighbourhood of $\hbar=0$.

To reduce two dimensions in (\ref{ME in c4}) we impose the symmetry in two orthogonal directions
$(\partial_{w}-\partial_{\tilde{w}})\Theta=0$ and
$(\partial_{z}-\partial_{\tilde{z}})\Theta=0$. This leads to 
\be
\label{Moyal H-P}
\partial^{2}_{w}\Theta +\partial^{2}_{z}\Theta +
\{ \partial_{w}\Theta\, ,\,\partial_{z}\Theta\}=0.
\ee
for $\Theta=\Theta(t,\hbar,w+\tilde{w}, z+\tilde{z},x^{1},x^{2})$. 

Substitute $w+\tilde{w}\mapsto w$ and $z+\tilde{z}\mapsto z$ and define 
$\theta:=\Theta_{0}(t,w, z, x^{1}, x^{2})$. From (\ref{analytic Theta}) one gets 
$\theta=\lim_{\hbar\rightarrow 0}\Theta$. Then, from the definition of deformation
quantization,  the bracket 
$\{\partial_{w}\Theta,\partial_{z}\Theta\}$ turns into Poisson bracket 
$\{\partial_{w}\theta,\partial_{z}\theta\}_{\mbox{\scriptsize{Poisson}}}$ in the 
classical limit $\hbar\rightarrow 0$. In this limit the equation (\ref{Moyal H-P}) leads to the 
\it Husain-Park \rm (HP) \it heavenly equation \rm (Park 1992, Husain 1994)
\be
\label{H-P}
\partial^{2}_{w}\theta +\partial^{2}_{z}\theta +
\{ \partial_{w}\theta\, ,\,\partial_{z}\theta\}_{\mbox{\tiny{Poisson}}}=0
\ee
(Comparing ME (\ref{ME in c4}) with (\ref{H-P}) we conclude that ME can be 
thought of as a quantum deformation of HP heavenly equation lifted to six 
dimensions.) 

The solution $\theta(t, w, z, x^{1},x^{2})$ of (\ref{H-P}) defines the heavenly metric
\bea\label{H-P metric}
ds^{2} &=& dw\cdot(\partial_{x^{1}}\partial_{w}\theta dx^{1}+\partial_{x^{2}}\partial_{w}\theta dx^{2})+
dz\cdot(\partial_{x^{1}}\partial_{z}\theta dx^{1}+\partial_{x^{2}}\partial_{z}\theta dx^{2})+\\
& & -\frac{1}{\{\partial_{w}\theta ,\partial_{z}\theta\}_{ \mbox{\tiny{Poisson}}}}
[(\partial_{x^{1}}\partial_{w}\theta dx^{1}+\partial_{x^{2}}\partial_{w}\theta dx^{2})^{2}+
(\partial_{x^{1}}\partial_{z}\theta dx^{1}+\partial_{x^{2}}\partial_{z}\theta dx^{2})^{2}].\nonumber
\eea
\brem\label{deformed H-P} \rm 

\mbox{ }

\bi\item On the other hand one can consider the (formal) deformation quantization of (\ref{H-P}). 
This leads to the equation (\ref{Moyal H-P}) but written
in the algebra of formal power series i.e.
\bdm
\Theta=\sum^{\infty}_{k=0}\hbar^{k}
\Theta_{k}(w,z,x^{1},x^{2})
\edm
This algebra is in general different from the algebra ${\cal A}$ and from the analytic algebra 
considered above. The series may not be convergent, but the reduction to the HP equation 
exists. This reduction  corresponds to the classical limit and in this limit any $\ast$-product
and any algebra of quantum observables reduce to the algebra of functions with Poisson bracket.
\ei\erem
\bexam \rm
In order to find a solution of (\ref{Moyal H-P}) consider the Cauchy data
\beas
\Theta|_{z=0}&=&\frac{\pi}{2}\cos(x^{1}+x^{2})-w\sin x^{2}\\
\partial_{z}\Theta|_{z=0}&=&-\sin x^{1}
\eeas
One easily computes from (\ref{Moyal H-P}) that for $k\geq 2$
\beas
\begin{array}{rcl}
\frac{\partial^{k}\Theta}{\partial z^{k}}|_{\mbox{\tiny{$z=0$}}}&=&
\{...\{\partial_{z}\Theta,\underbrace{\partial_{w}\Theta\},...,
\partial_{w}\Theta\}}_{\mbox{\tiny{$k-1$ times}}}|_{\mbox{\tiny{$z=0$}}}\\
&=&-(\frac{2}{\hbar}\sin\frac{\hbar}{2})^{k-1}\cos^{k-1}x^{2}\frac{d^{k-2}}{d(x^{1})^{k-2}}\cos x^{1}
\end{array}\eeas
The Cauchy-Kowalewska form of the solution 
$\Theta=\sum\frac{1}{k!}\frac{\partial^{k}\Theta}{\partial z^{k}}|_{\mbox{\tiny{$z=0$}}}z^{k}$ reads
\bea
\label{example solution in series form}
\Theta&=&\Theta|_{z=0}+\partial_{z}\Theta|_{z=0}z+
\sum_{m=1}^{\infty}\frac{(-1)^{m}}{2m!}
(\frac{2}{\hbar}\sin\frac{\hbar}{2}\cos x^{2})^{2m-1}(\cos x^{1})\,z^{2m}\nonumber\\
& &+\sum_{m=1}^{\infty}\frac{(-1)^{m-1}}{(2m+1)!}
(\frac{2}{\hbar}\sin\frac{\hbar}{2}\cos x^{2})^{2m}(\sin x^{1})\,z^{2m+1}
\eea
%
Finally, the solution of (\ref{Moyal H-P}) is given by
\be
\label{solution}
\Theta=\frac{\pi}{2}\cos(x^{1}+x^{2})-w\sin x^{2}+
\frac{\cos [z\cdot\frac{2}{\hbar}\sin\frac{\hbar}{2}\cdot\cos x^{2}+x^{1}]-
\cos x^{1}}{\frac{2}{\hbar}\sin\frac{\hbar}{2}\cos x^{2}}
\ee
This is an analytic function of $\hbar$. The absence of the parameter $t$ may be thought of 
as a consequence of a suitable choice of coordinates $w$ and $z$.

To simplify the notation we denote $p\equiv x^{1}$ and $q\equiv x^{2}$. Thus 
the solution $\Theta=\Theta(\hbar,w,z,p,q)$ and (\ref{solution}) reads
\beas
\Theta&=&\frac{\pi}{2}\cos(p+q)-w\sin q+
\frac{\cos [z\cdot\frac{2}{\hbar}\sin\frac{\hbar}{2}\cdot\cos q+p]-
\cos p}{\frac{2}{\hbar}\sin\frac{\hbar}{2}\cos q}\\
&=&\frac{\pi}{2}\cos(p+q)-w\sin q-
\int_{0}^{ z}\sin (\zeta\frac{2}{\hbar}\sin\frac{\hbar}{2}\cdot\cos q+p)d\zeta
\eeas
In the limit $\hbar\rightarrow 0$ one gets the solution of (\ref{H-P})
\beas
\theta &=&\frac{\pi}{2}\cos(p+q)-w\sin q+
\frac{\cos [z\cdot\cos q+p]-
\cos p}{\cos q}\\
&=&\frac{\pi}{2}\cos(p+q)-w\sin q-
\int_{0}^{ z}\sin (\zeta\cdot\cos q+p)d\zeta
\eeas
Substituting this $\theta$ into (\ref{H-P metric}) we obtain the following heavenly metric
\bea\label{example metric}
ds^{2}&=&-\cos q\,dw\,dq+\cos(z\cos q+p)\cdot (z\sin q\,dq-dp)dz\\
& & - \frac{1}{\cos q\cos(z\cos q+p)}[\cos^{2}q\, dq^{2}+\cos^{2}(z\cos q+p)\cdot
(z\sin q\, dq-dp)^{2}]\nonumber
\eea
In order to classify the heavenly metric (\ref{example metric}), we rewrite it in the null tetrad form \h
\mbox{$ds^{2}=2\mbox{\boldmath{$e$}}^{1}\mbox{\boldmath{$e$}}^{2}+
2\mbox{\boldmath{$e$}}^{3}\mbox{\boldmath{$e$}}^{4}$.}\h 
In what follows we use the convention of Pleba\'{n}ski (1975), Pleba\'{n}ski and Przanowski (1988).
One quickly finds that the 1-forms
\beas
\mbox{\boldmath{$e$}}^{1}&=&\frac{1}{\sqrt{2}}\Phi^{-1}[\cos q\,dz-(z\sin q\,dq-dp)]\\
\mbox{\boldmath{$e$}}^{2}&=&\frac{1}{\sqrt{2}}(z\sin q\, dq-dp) \\
\mbox{\boldmath{$e$}}^{3}&=&-\frac{1}{\sqrt{2}}dq \\
\mbox{\boldmath{$e$}}^{4}&=&\frac{1}{\sqrt{2}}[\cos q\,dw+\Phi\,dq]
\eeas
where $\Phi:=\frac{\cos q}{\cos (z\cos q+p)}$, define a null tetrad for the metric (\ref{example metric}).
Then
\beas
\begin{array}{rcccrcl}
dq &=& -\sqrt{2}\mbox{\boldmath{$e$}}^{3} &,&
dw &=& \frac{\sqrt{2}}{\cos q}(\mbox{\boldmath{$e$}}^{4}+\Phi\,\mbox{\boldmath{$e$}}^{3})\\
dp &=& -\sqrt{2}(\mbox{\boldmath{$e$}}^{2}+z\sin q\mbox{\boldmath{$e$}}^{3}) &,&
dz &=& \frac{\sqrt{2}}{\cos q}(\mbox{\boldmath{$e$}}^{2}+\Phi\,\mbox{\boldmath{$e$}}^{1})
\end{array}
\eeas

The  first Cartan structure equations  
$d\mbox{\boldmath{$e$}}^{a}=-\Gamma^{a}_{\mbox{ }b}\wedge\mbox{\boldmath{$e$}}^{b}$
read
\beas
d\mbox{\boldmath{$e$}}^{1}&=&-\sqrt{2}\,\mbox{tg}\;q\,\mbox{\boldmath{$e$}}^{3}\wedge\mbox{\boldmath{$e$}}^{1}\\
d\mbox{\boldmath{$e$}}^{2}&=&-\sqrt{2}\,\mbox{tg}\;q\,[
\mbox{\boldmath{$e$}}^{2}\wedge\mbox{\boldmath{$e$}}^{3}
+\Phi\,\mbox{\boldmath{$e$}}^{1}\wedge\mbox{\boldmath{$e$}}^{3}]\\
d\mbox{\boldmath{$e$}}^{3}&=& 0\\
d\mbox{\boldmath{$e$}}^{4}&=&\sqrt{2}\;[\;\mbox{tg}\; q
\mbox{\boldmath{$e$}}^{3}\wedge\mbox{\boldmath{$e$}}^{4}+\Phi^{2}\mbox{tg}\;(z\cos q+p)
\mbox{\boldmath{$e$}}^{3}\wedge\mbox{\boldmath{$e$}}^{1}\;]
\eeas
Consequently, the only non zero connection 1-forms are
\beas
\Gamma_{12}&=&-\sqrt{2}\mbox{tg}\,q\, \mbox{\boldmath{$e$}}^{3}\\
\Gamma_{31}&=&-\sqrt{2}\Phi\,[\mbox{tg}\,q\, \mbox{\boldmath{$e$}}^{1}\,+\,
\Phi\,\mbox{tg}\,(z\cos q+p)\,\mbox{\boldmath{$e$}}^{3}]\\
\Gamma_{34}&=&-\sqrt{2}\mbox{tg}\,q\, \mbox{\boldmath{$e$}}^{3}
\eeas
First note that the  \it dotted \rm spinor connection 
$\Gamma_{\stackrel{\cdot}{A}\stackrel{\cdot}{B}}$ vanishes 
(Pleba\'{n}ski 1975, Pleba\'{n}ski and Przanowski 1988) since
\bdm
\Gamma_{41}=\frac{1}{2}(-\Gamma_{12}+\Gamma_{34})=\Gamma_{32}=0\h\h
\Rightarrow\h\h\Gamma_{\stackrel{\cdot}{A}\stackrel{\cdot}{B}}=0.
\edm
Thus the curvature form is self-dual and indeed, the metric $ds^{2}$ given by
(\ref{example metric}) describes the heavenly space.

On the other hand the only nontrivial second Cartan structure equation for the \it undotted 
spinor connection \rm  is
\bdm
d\Gamma_{31}+(\Gamma_{12}+\Gamma_{34})\wedge\Gamma_{31}=\frac{1}{2}C^{(1)}
\mbox{\boldmath{$e$}}^{3}\wedge\mbox{\boldmath{$e$}}^{1}.
\edm
where $\frac{1}{2}C^{(1)}=C_{2222}$ 
and $C_{ABCD}$ is \it undotted Weyl spinor \rm i.e. the spinor image of the self-dual part of 
the Weyl tensor.

This gives
\bdm
C^{(1)}=4\Phi\,[\frac{1+2\sin^{2}q}{\cos^{2}q}\,+\,
\frac{1+2\sin^{2}(z\cos q+p)}{\cos^{2}(z\cos q+p)}\Phi^{2}\,].
\edm
Thus the only nonvanishing component of the Weyl spinor 
is $C^{(1)}$, what means that the heavenly space $\M$ is of the type
$[4]\times[-]$ ($N\times 0$).

Summarizing, the solution (\ref{solution}) leads in a classical limit $\hbar\rightarrow 0$
to a heavenly space of the type $[4]\times [-]$ described by the metric (\ref{example metric}).
This metric is complex if $\{w,z,p,q\}$ are complex or real of signature $(++ --)$ if 
$\{w,z,p,q\}$ are real.

As has been pointed out by Maciej Dunajski the transformation $u=\sin q$, $v=\sin(z\cos q+p)$
brings the metric (\ref{example metric}) to a simple pp-wave (Pleba\'{n}ski 1975)
form
\bdm
ds^{2}=-\,dw\,du+dv\,dz - \frac{1}{\sqrt{(1-u^{2})(1-v^{2})}}( du^{2}+dv^{2})
\edm
\eexam
%
%
\setcounter{equation}{0}
\section{Reduction of ME to the $sl(N,\mbox{\boldmath{$C$}})$ SDYM equation}
\label{Reduction of ME to $su(N)$ SDYM system}
In this section we  examine the conditions under which the solution of ME
(\ref{ME in H}) defines an analytic solution of $sl(N,\mbox{\boldmath{$C$}})$ SDYM equation.

As has been pointed out in the remark \ref{deformed H-P} of the previous section, the reduction of 
ME to HP heavenly equation is in a sense trivial. This  is due to the fact that 
Poisson algebra (the algebra of Hamiltonian vector fields over $(\Sigma^{2},\mbox{\boldmath{$\omega$}})$)  
by the very definition can be embedded into any deformed algebra.

In the case of reduction to $sl(N,\mbox{\boldmath{$C$}})$ algebra one has to find an
algebra $({\cal A},\ast)$ for which there exists a Lie algebra homomorphism 
${\cal A}\rightarrow sl(N,\mbox{\boldmath{$C$}})$.

The noncommutative $\ast$-product exists on each symplectic space $(\Sigma^{2n},\mbox{\boldmath{$\omega$}})$. 
This was proved by De Wilde and Lacomte (1992) and  Fedosov (1994, 1996). In this paper we 
restrict ourselves to two dimensional torus $T^{2}$ with coordinates $(x^{1},x^{2})$ and 
symplectic form $\mbox{\boldmath{$\omega$}}=dx^{1}\wedge dx^{2}$. Fedosov's $\ast$ 
multiplication on $T^{2}$  can be chosen to be just the Moyal $\ast$-product (Fedosov 1994, 1996).
\bdm
\forall f,g\in C^{\omega}(T^{2})\h\h\h 
f\ast g:=\sum_{k=0}^{\infty}\frac{1}{k!}(\frac{i\hbar}{2})^{k}
\omega^{i_{1}j_{1}}...\omega^{i_{k}j_{k}}
\frac{\partial^{k}f}{\partial x^{i_{1}}...\partial x^{i_{k}}}
\frac{\partial^{k}g}{\partial x^{j_{1}}...\partial x^{j_{k}}}.
\edm

Each analytic function $f$ on $T^{2}$, $f\in C^{\omega}(T^{2})$ can be expanded into 
Fourier series with respect to the basis 
\be
E_{\vec{m}}=\exp[i(m_{1}x^{1}+m_2 x^{2})]\, ,\h\h\h\h
\vec{m}:=(m_{1}, m_2)\in\mbox{\boldmath{$Z$}}\times\mbox{\boldmath{$Z$}}.
\ee
In this basis the Moyal $\ast$-product has a simple form
\be
\label{ast on T2}
E_{\vec{m}}\ast E_{\vec{n}}=
\exp(\frac{i\hbar}{2}\vec{m}\times\vec{n})E_{\vec{m}+\vec{n}}\, ;\h\h\h
\vec{m}\times\vec{n}=m_{1}n_{2}-m_{2}n_{1}.
\ee
Consequently the Moyal bracket
\be
\label{bracket on T2}
\{E_{\vec{m}},E_{\vec{n}}\}=\frac{2}{\hbar}\sin(\frac{\hbar}{2}\vec{m}\times\vec{n})
E_{\vec{m}+\vec{n}}.
\ee
\brem \rm

\mbox{ }

\bi\item The formulas (\ref{ast on T2}) and (\ref{bracket on T2}) prove that the Moyal
$\ast$-product and Moyal bracket are analytic in deformation parameter $\hbar$.

This feature is fundamental to find a reduction of the formal $\ast$-algebra to a finite 
dimensional algebra. It makes possible to put a particular value of $\hbar$ 
and then to come from formal power series to analytic functions. Then in this case one 
can expect that the solutions of the $\ast$-SDYM  equation analytic in $\hbar$ define
a family of solutions to $sl(N,\mbox{\boldmath{$C$}})$ SDYM equations.
\ei\erem
In order to find a relation between the Lie algebras $(C^{\omega}(T^{2}),\{\cdot,\cdot\})$ and 
$sl(N,\mbox{\boldmath{$C$}})$ we have to introduce a suitable basis for this latter 
algebra.
\subsection{Trygonometric structure constants for $sl(N,\mbox{\boldmath{$C$}})$}
Here we briefly summarize the results of (Fairlie et al 1990). In that distinguished work the basis for finite 
diemensional Lie algebras was constructed in which the structure constants are trygonometric functions.

For any $\vec{m}:=(m_{1},m_{2})\in\mbox{\boldmath{$Z$}}\times\mbox{\boldmath{$Z$}}$, 
where $\mbox{\boldmath{$Z$}}$ is a group of rational numbers, consider an $N\times N$ matrix
\be
L_{\vec{m}} := \frac{iN}{2\pi} \omega^{\frac{m_{1}m_{2}}{2}} S^{m_{1}}T^{m_{2}}.
\ee
where \small
\beas
S:=\sqrt{\omega}
\left(
\begin{array}{ccccc}
1&0&0&\ldots&0\\
0&\omega&0&\ldots&0\\
0&0&\omega^{2}&\ldots&0\\
.&.&.&\ldots&.\\
0&0&0&\ldots&\omega^{N-1}
\end{array} \right)
\h , \h\h
T:=\left(
\begin{array}{ccccc}
0&1&0&\ldots&0\\
0&0&1&\ldots&0\\
.&.&.&.&.\\
0&0&0&\ldots&1\\
-1&0&.0&\ldots&0
\end{array} \right)
\eeas 
\normalsize
and $\omega$ is the $N$th root of $1$. We take $\omega :=exp(\frac{2\pi i}{N})$. 

The unitary matrices $S$ and $T$ satisfy: $T\, S=\omega S\, T$, 
and $S^N=T^N=-\mbox{\boldmath{$1$}}$.

The system of $(N^{2}-1)$ matrices $\{L_{\vec{\mu}}\}$, $0\leq \mu_{1}< N$, $0\leq \mu_{2}< N$ 
apart from $(\mu_{1},\mu_{2})=(0,0)$, is linearly independent.
\bproposition\label{properties of L}  Matrices $L_{\vec{m}}$ have the following properties \rm

\benum
\item $L_{\vec{m}+N\vec{r}}= (-1)^{(m_1+1)r_2+(m_2+1)r_1+Nr_1 r_2} L_{\vec{m}}$
\item $\mbox{Tr}L_{\vec{m}}=0$ \h\h\h \mbox{apart from} \h\h\h $m_1 = m_2 = 0$ 
{\mbox{\scriptsize{modulo}}}$\,N$
\item $Tr L_{N\vec{r}}:= (-1)^{r_2+r_1+Nr_1 r_2}\,\frac{iN^2}{2\pi}$.
\item $L_{\vec{m}}L_{\vec{n}}=\frac{iN}{2\pi}
\omega^{\frac{\vec{n}\times \vec{m}}{2}} L_{\vec{m}+\vec{n}}$; \h\h
$\vec{n}\times\vec{m}:=n_{1} m_{2}-n_{2} m_{1}$
\item $L_{\vec{m}}^{\dagger}=-L_{-\vec{m}}=(\frac{N}{2\pi})^2L_{\vec{m}}^{-1}$
\item $\det L_{\vec{m}}:= (-1)^{N(m_1+m_2+m_1 m_2)}\,(\frac{iN}{2\pi})^N$ \h\h $\Box$
\eenum \eproposition
\bcorollary 
\mbox{                                                                                    }
\benum 
\item The matrices $L_{\vec{\mu}}$, $0\leq \mu_{1}< N$, $0\leq \mu_{2}< N$ 
and $\vec{\mu}\neq (0,0)$ are traceless and linearly independent. Therefore, they 
constitute  a basis of the $sl(N,\mbox{\boldmath{$C$}})$ algebra. 
For this basis  the structure constants are 
trygonometric functions
\be
\label{5.2}
[L_{\vec{\mu}},L_{\vec{\nu}}]=\frac{N}{\pi}sin(\frac{\pi}{N}\vec{\mu}\times\vec{\nu})
L_{\vec{\mu}+\vec{\nu}}
\ee
\item Appropriate linear combinations of $L_{\vec{\mu}}$ define anti-hermitian matrices which 
constitute a basis for $su(N)$ algebra, for arbitrary $N\geq 2$. \h\h\h $\Box$
\eenum 

\ecorollary
\brem\label{mu and nu}\rm

\mbox{                                                                                 }

\bi\item In 
all the paper 
the Greek indecies $\vec{\mu},\vec{\nu},...$ etc. satisfy 
$\vec{\mu}:=(\mu_{1},\mu_{2})\neq (0,0)$ and $0\leq \mu_{1}\leq N-1$, $0\leq \mu_{2}\leq N-1$.
\ei\erem
%
                 
\subsection{From the solution of ME to the solution of the $sl(N,\mbox{\boldmath{$C$}})$ SDYM equation}
As has been said at the beginning of this section, the Moyal bracket 
$\{E_{\vec{m}},E_{\vec{n}}\}=\frac{2}{\hbar}\sin(\frac{\hbar}{2}\vec{m}\times\vec{n})
E_{\vec{m}+\vec{n}}$ is an analytic function of the deformation parameter. 
Thus one can replace the formal power series by convergent ones and put a particular value 
for $\hbar$. Taking $\hbar=\frac{2\pi}{N}$ one gets
\be
\label{nawias dla 2pi/N}
\{E_{\vec{m}},E_{\vec{n}}\}=\frac{N}{\pi}\sin(\frac{\pi}{N}\vec{m}\times\vec{n})
E_{\vec{m}+\vec{n}}.
\ee
The structure constants are the same as the structure constants of $sl(N,\mbox{\boldmath{$C$}})$ 
algebra in the basis $L_{\vec{\mu}}$. For this reason, for each $N\geq 2$, we can define a 
 Lie algebra homomorphism  
$\chi_{\mbox{\tiny{$N$}}}:Span\{E_{\vec{m}}\}\stackrel{\mbox{\scriptsize{onto}}}{\longrightarrow} 
sl(N,\mbox{\boldmath{$C$}})$, 
by
\bea
\label{representation}
\chi_{\mbox{\tiny{$N$}}}: \left\{ \begin{array}{ccc}
E_{\vec{\mu}+N\vec{r}}&\longmapsto &(-1)^{(\mu_1+1)r_2+(\mu_2+1)r_1+Nr_1 r_2} L_{\vec{\mu}}\\
E_{N\vec{r}}&\longmapsto &0
\end{array}\right.
\eea
where $\vec{\mu}$ has been defined in the remark \ref{mu and nu} and 
$\vec{r}:=(r_1,r_2)\in\mbox{\boldmath{$Z$}}\times\mbox{\boldmath{$Z$}}$. 

\vspace{1 ex}

Recall, that in the  $K$-Newman formalism
the $sl(N,\mbox{\boldmath{$C$}})$ SDYM system on the heavenly background  
can be reduced to one equation
\be
\label{SDYM sl(N,C)}
g^{\tilde{\beta}\alpha}\partial_{\alpha}\partial_{\tilde{\beta}}\vartheta +
\frac{1}{2G}\epsilon^{\alpha\beta}[\partial_{\alpha}\vartheta\, ,
\, \partial_{\beta}\vartheta ]=0\, ;\h\h\h
\vartheta(y,\tilde{y},z,\tilde{z})\in sl(N,\mbox{\boldmath{$C$}})
\ee
(Newman 1978, Leznov 1988, Parkes 1992, Pleba\'{n}ski and Przanowski 1996, Mason
and Woodhouse 1996, Przanowski and Forma\'{n}ski 1999, Forma\'{n}ski 2004).

The following theorem gives the relations between ME (\ref{ME in H}) and 
$sl(N,\mbox{\boldmath{$C$}})$ SDYM equation (\ref{SDYM sl(N,C)}) (see 
also Przanowski and Forma\'{n}ski (1999)).
\bthrm\label{o homomorfizmie rozwiazan} 
Each analytic solution of $sl(N,\mbox{\boldmath{$C$}})$ SDYM 
equation (\ref{SDYM sl(N,C)}) is an image in the homomorphism $\chi_{\mbox{\tiny{$N$}}}$ of 
some solution of ME (\ref{ME in H}).
\ethrm
\bf Proof\rm . Suppose that $\vartheta$ satisfies  the equation (\ref{SDYM sl(N,C)}).
Without any loss of generality we can take such coordinates on $\M$ that $g_{z\tilde{z}}\neq 0$ i.e. 
$g^{\tilde{w}w}\neq 0$. After the change of coordinates
\be
\label{zmienne y}
w=\frac{1}{2}(y+\tilde{y})\, ,\h\h \tilde{w}=\frac{1}{2}(y-\tilde{y})\h\h \Rightarrow\h\h
y=w+\tilde{w}\, ,\h\h \tilde{y}=w-\tilde{w}
\ee
the $sl(N,\mbox{\boldmath{$C$}})$ SDYM equations (\ref{SDYM sl(N,C)}) reads

\bea
\label{SDYM-Kowalewska}
\partial^{2}_{y}\vartheta &=&\partial^{2}_{\tilde{y}}\vartheta
-\frac{1}{g^{\tilde{w}w}}(\,
g^{\tilde{z}z}\partial_{z}\partial_{\tilde{z}}\vartheta+
g^{\tilde{z}w}\partial_{y}\partial_{\tilde{z}}\vartheta+
g^{\tilde{z}w}\partial_{\tilde{y}}\partial_{\tilde{z}}\vartheta+
g^{\tilde{w}z}\partial_{z}\partial_{y}\vartheta-
g^{\tilde{w}z}\partial_{z}\partial_{\tilde{y}}\vartheta\, )\nonumber\\ 
& &-\frac{1}{G\,g^{\tilde{w}w}}\,
[\partial_{y}\vartheta+\partial_{\tilde{y}}\vartheta\, ,\, \partial_{z}\vartheta]\,
\eea
Each analytic solution of the above equation can be obtained by the Cauchy-Kowalewska method 
for some Cauchy data. Assume the Cauchy data of the form
\bea
\label{Cauchy data}
\vartheta|_{y=0}&=&\sum_{\vec{\mu}}\vartheta^{(0)}_{\vec{\mu}}(\tilde{y},z,\tilde{z})L_{\vec{\mu}}
\nonumber\\
\partial_{y}\vartheta|_{y=0}&=&\sum_{\vec{\mu}}\vartheta^{(1)}_{\vec{\mu}}(\tilde{y},z,\tilde{z})
L_{\vec{\mu}}
\eea
(Remember that the index $\vec{\mu}$ in above sums satisfies the conditions of remark \ref{mu and nu}).

Consider ME. Under the same assumptions about $\M$ it takes the same form as 
(\ref{SDYM-Kowalewska}) 
\bea
\label{ME-Kowalewska}
\partial^{2}_{y}\Theta &=&\partial^{2}_{\tilde{y}}\Theta-\frac{1}{g^{\tilde{w}w}}(\,
g^{\tilde{z}z}\partial_{z}\partial_{\tilde{z}}\Theta+
g^{\tilde{z}w}\partial_{y}\partial_{\tilde{z}}\Theta+
g^{\tilde{z}w}\partial_{\tilde{y}}\partial_{\tilde{z}}\Theta+
g^{\tilde{w}z}\partial_{z}\partial_{y}\Theta-
g^{\tilde{w}z}\partial_{z}\partial_{\tilde{y}}\Theta)\nonumber\\ & &
-\frac{1}{G\,g^{\tilde{w}w}}\{\partial_{y}\Theta+\partial_{\tilde{y}}\Theta\, ,\, \partial_{z}\Theta\}\,)
\eea
One can seek for the solution of (\ref{ME-Kowalewska}) for which the Cauchy data as projected by
$\chi_{\mbox{\tiny{$N$}}}$ give (\ref{Cauchy data}) i.e.
\bea
\label{war poczat na Theta}
\Theta|_{y=0}&=&\sum_{\vec{\mu}}\vartheta^{(0)}_{\vec{\mu}}(\tilde{y},z,\tilde{z})E_{\vec{\mu}}
\nonumber\\
\partial_{y}\Theta|_{y=0}&=&\sum_{\vec{\mu}}\vartheta^{(1)}_{\vec{\mu}}(\tilde{y},z,\tilde{z})
E_{\vec{\mu}}
\eea

Thus the  Cauchy-Kowalewska method in both cases gives the solution in a form of a power 
series with respect to the coordinates $\{y,\tilde{y},z,\tilde{z}\}$.  For simplicity we write here
\bea
\label{rozwiazanie sl(N,C)}
\vartheta &=&\sum_{j=0}^{\infty}
\,\frac{1}{j!}(\sum_{\vec{\mu}}\,\vartheta^{(j)}_{\vec{\mu}}(\tilde{y},z,\tilde{z})\,L_{\vec{\mu}})\, y^{j}
\hspace{12 ex}\mbox{in}\hspace{2 ex}sl(N,\mbox{\boldmath{$C$}}).\\
\label{rozwiazanie ME}
\Theta &= &\lim_{K\rightarrow\infty}\sum_{j=0}^{K}\frac{1}{j!}\,(
\sum_{\vec{n}=\vec{0}}^{\vec{M}_{j}}\Theta^{(j)}_{\vec{n}}(\hbar;\tilde{y},z,\tilde{z})
E_{\vec{n}}\, )y^{j}\,
\hspace{5 ex}\mbox{in}\hspace{2 ex}C^{\omega}(T^{2}).
\eea
In the last sum the symbol $\sum_{\vec{n}=\vec{0}}^{\vec{M}_{j}}$ means sum over 
$0\leq n_{1}\leq M_{1j}$, $0\leq n_{2}\leq M_{2j}$, where for each $j$ the numbers 
$M_{1j}$ and $M_{2j}$ are finite.

\noindent At each order $K$ of the approximation we have
\bdm
\chi_{\mbox{\tiny{$N$}}}(\sum_{j=0}^{K}\frac{1}{j!}\,(
\sum_{\vec{n}=\vec{0}}^{\vec{N}_{j}}\Theta^{(j)}_{\vec{n}}(\frac{2\pi}{N};\tilde{y},z,\tilde{z})
E_{\vec{n}}\, y^{j}\,))=\sum_{\vec{\mu}}\sum_{j=0}^{K}
(\,\frac{1}{j!}\,\vartheta^{(j)}_{\vec{\mu}}(\tilde{y},z,\tilde{z})\, y^{j}\,)L_{\vec{\mu}}
\edm
Thus the solution (\ref{rozwiazanie sl(N,C)}) is defined by the solution of 
(\ref{rozwiazanie ME}) and it can be called an image 
of the solution (\ref{rozwiazanie ME}) in $\chi_{\mbox{\tiny{$N$}}}$.

\vspace{0.5 ex}

\noindent This completes the proof. $\Box$

\vspace{1 ex}

The above theorem says that the SDYM equation in $sl(N,\mbox{\boldmath{$C$}})$ algebra 
for arbitrary $N\geq 2$ is a reduction of ME. As is known from the works (Mason and 
Sparling 1989, Chakravarty and Ablowitz 1992, Tafel 1993) the  
$sl(N,\mbox{\boldmath{$C$}})$ SDYM equations reduce to many integrable systems. 
Thus the theorem \ref{o homomorfizmie rozwiazan} finally embeds those integrable 
equations in ME. In a sense this justyfies the name \it master equation \rm for 
the equation (\ref{ME in H}).

\vspace{1 ex}

The way in which the theorem \ref{o homomorfizmie rozwiazan} has been proved
suggests a method to obtain analytic 
solutions of $sl((N,\mbox{\boldmath{$C$}})$ SDYM equation (\ref{SDYM sl(N,C)})
from some solutions of ME. Given a solution 
$\Theta(t,\hbar,y,\tilde{y},z,\tilde{z},x^{1},x^{2}$) of (\ref{ME-Kowalewska}) 
analytic in all coordinates and in t and $\hbar$ as well for which the Cauchy data read 
\bea
\label{Cauchy data for Theta}
\Theta|_{y=0}&=&\sum_{\vec{m}=-\vec{M}}^{\vec{M}}
\Theta^{(0)}_{\vec{m}}(t,\hbar,\tilde{y},z,\tilde{z})E_{\vec{m}}
\nonumber\\
\partial_{y}\Theta|_{y=0}&=&\sum_{\vec{m}=-\vec{M}}^{\vec{M}}
\Theta^{(1)}_{\vec{\mu}}(t,\hbar,\tilde{y},z,\tilde{z})
E_{\vec{m}}
\eea
by applaying the homomorphism $\chi_{\mbox{\tiny{$N$}}}$ we get 
\bea
\label{Cauchy data for sl(N,C)}
\chi_{\mbox{\tiny{$N$}}}(\Theta|_{y=0})&=&\sum_{\vec{m}=-\vec{M}}^{\vec{M}}
\Theta^{(0)}_{\vec{m}}(t,\frac{2\pi}{N},\tilde{y},z,\tilde{z})L_{\vec{m}}=
\sum_{\vec{\mu}}\vartheta^{(0)}_{\vec{\mu}}(t,\tilde{y},z,\tilde{z})L_{\vec{\mu}}
\nonumber\\
\lefteqn{\chi_{\mbox{\tiny{$N$}}}(\partial_{y}\Theta|_{y=0})=}\\
&=&\sum_{\vec{m}=-\vec{M}}^{\vec{M}}
\Theta^{(1)}_{\vec{\mu}}(t,\frac{2\pi}{N},\tilde{y},z,\tilde{z})
L_{\vec{m}}=\sum_{\vec{\mu}}\vartheta^{(1)}_{\vec{\mu}}(t,\tilde{y},z,\tilde{z})L_{\vec{\mu}}\nonumber
\eea
where the index $\vec{\mu}$ runs through $(0,1),(1,0),...,(N-1,N-1)$ according to the 
remark \ref{mu and nu}. 

If one considers (\ref{Cauchy data for sl(N,C)}) to be the Cauchy data for 
$sl((N,\mbox{\boldmath{$C$}})$ SDYM equation written in the form (\ref{SDYM-Kowalewska})
then one gets a respective  solution $\vartheta$ of this equation. One easily finds
that the $\vartheta$ can be obtained directly from $\Theta$ as follows: 

\noindent One expands $\Theta$ into Fourier series in the basis $E_{\vec{m}}$
\bdm
\Theta=\sum_{\vec{m}\in \mbox{\boldmath{$Z$}}\times\mbox{\boldmath{$Z$}}}
\Theta_{\vec{m}}(t,\hbar ,w,z,\tilde{w},\tilde{z})E_{\vec{m}}.
\edm 
Then one makes the substitutions $\hbar\mapsto\frac{2\pi}{N}$ and 
$E_{\vec{m}}\mapsto L_{\vec{m}}$. At this stage 
a formal series in $L_{\vec{m}}$ is obtained, for which one applies the property 1 
of the \mbox{proposition \ref{properties of L}} 
and gathers the elements standing at the same vectors $L_{\vec{\mu}}$

\noindent Consequently, one gets 
\be
\label{chi od Theta}
\vartheta=\sum_{\vec{\mu}}
(\,\sum_{\vec{r}\in \mbox{\boldmath{$Z$}}\times\mbox{\boldmath{$Z$}}}
(-1)^{(\mu_1+1)r_2+(\mu_2+1)r_1+Nr_1 r_2} \Theta_{\vec{\mu}+N\vec{r}}\, )
L_{\vec{\mu}}.
\ee
The above procedure will be used in the example 2 to construct a sequence 
of $su(N)$ chiral fields defined by the solution $\Theta$ of ME given in 
the example 1.
\setcounter{equation}{0}
\section{From chiral fields to heavenly spaces}
\label{From chiral fields to heavenly space}
Once again consider the reduced form (\ref{Moyal H-P}) of ME discussed in Section 
\ref{Reduction of ME to heavenly equation}. Let the symplectic manifold be the same as in  
section \ref{Reduction of ME to $su(N)$ SDYM system}, i.e. the two dimensional 
torus $T^{2}$ with coordinates $x^{1}\equiv p$, $x^{2}\equiv q$ and symplectic form 
\mbox{$\mbox{\boldmath{$\omega$}}=dp\wedge dq$.} 
In the same way as is done for the SDYM equation (theorem \ref{o homomorfizmie rozwiazan}) the homomorphism
$\chi_{\mbox{\tiny{N}}}$ projects the solution $\Theta$ of (\ref{Moyal H-P}) onto the solution
$\vartheta=\chi_{\mbox{\tiny{N}}}(\Theta)$: $\M^{2}\rightarrow sl(N,\mbox{\boldmath{$C$}})$, 
$N=2,3,...$ of the following equation
\be
\label{chiral}
\partial^{2}_{w}\vartheta +\partial^{2}_{z}\vartheta +
[ \partial_{w}\vartheta\, ,\,\partial_{z}\vartheta]=0.
\ee
This equation describes the chiral field, i.e. harmonic map 
\mbox{$g:\M^{2}\rightarrow SL(N,\mbox{\boldmath{$C$}})$.} 

Indeed, each harmonic map on two dimensional
manifold $\M^{2}$ with coordinates $\{w,z\}$ and diagonal metric 
$\eta_{\alpha\beta}=\mbox{diag}(+,+)$ fulfils the chiral equation
\be
\label{chiral2}
\partial_{w}(g^{-1}\partial_{w}g)+\partial_{z}(g^{-1}\partial_{z}g)=0 .
\ee
If one denotes $A_{\alpha}=g^{-1}\partial_{\alpha}g$, $\alpha\in\{w,z\}$, then this 
substitution implies the integrability condition 
 $\partial_{w}A_{z}-\partial_{z}A_{w}+[A_{w},A_{z}]=0$. Thus the chiral equation takes
the form of the system
\bea
\label{chiral3}
\begin{array}{cl}
&\partial_{w}A_{z}-\partial_{z}A_{w}+[A_{w},A_{z}]= 0\\
&\partial_{w}A_{w}+\partial_{z}A_{z}=0.
\end{array}
\eea
From the second equations of (\ref{chiral3}) one infers that there exists 
\mbox{$\vartheta: \M^{2}\rightarrow sl(N,\mbox{\boldmath{$C$}})$} such that 
\mbox{$A_{w}=-\partial_{z}\vartheta$} and $A_{z}=\partial_{w}\vartheta$. 
Substituting these $A_{w}$ and $A_{z}$ into the first equation of (\ref{chiral3}) we get
(\ref{chiral}).

In the present case theorem \ref{o homomorfizmie rozwiazan} of the previous section 
asserts that given a suitable solution $\Theta$ for $\hbar=\frac{2\pi}{N}$ one obtains 
the sequence of chiral fields $\vartheta=\chi_{\mbox{\tiny{N}}}(\Theta)$, $N=2,3,...$ 
satysfying (\ref{chiral}).
On the other hand, it is known that the solution $\Theta$ tends in a classical limit 
$\hbar\rightarrow 0$ to the solution $\theta$ of the HP heavenly equation (see section 
\ref{Reduction of ME to heavenly equation}). This makes the following diagram commutative

\setlength{\unitlength}{5ex}
\begin{picture}(9,5)
\put(2.7,3){$\chi_{\mbox{\tiny{N}}}$}
\put(4.8,4.4){$\Theta$}
\put(4.5,4){\vector(-1,-1){2}}
\put(1.5,1){$\vartheta$}
\put(2.5,1.3){\vector(1,0){4.8}}
\put(4.4,0.5){$N\rightarrow\infty$}
\put(7.7,1){$\theta$}
\put(6.6,3){$\hbar\rightarrow 0$}
\put(5.3,4){\vector(1,-1){2}}
\end{picture}

From this diagram one can conclude that the limiting process 
$\hbar \rightarrow 0$ is equivalent to  $N\rightarrow\infty$. It suggests that a 
better understanding of the $sl(\infty,\mbox{\boldmath{$C$}})$ algebra (see Ward 1992) 
can be achieved by lifting the Poisson algebra to the Moyal bracket algebra.
\brem\rm

\mbox{ }

\bi

\item The construction presented above allows one to interpret the heavenly equation 
as a chiral field equation on $\M^{2}$ with value in the algebra of hamiltonian vector 
fields on $T^{2}$. This algebra corresponds to the group of diffeomorphisms 
$SDiff(T^{2})$ which leave the volume form $\mbox{\boldmath{$\omega$}}$ unchanged.

\item This construction gives also a positive answer to the question asked by Ward (1990b): 
\it Can one construct a sequence of $SU(N)$ chiral fields, for $N=2,3,...$, tending to 
a curved space in the limit? \rm 
\ei\erem
In what follows we give an example of sequence of $su(N)$ chiral fields tending to 
a curved heavenly space for $N\rightarrow\infty$. (First, an example of sequence of
$sl(N,\mbox{\boldmath{$C$}})$ chiral fields definig a heavenly space in a limit
was found in (Przanowski et al 1998)).
\bexam\rm

\mbox{ }

In the example 1 we have found the solution $\Theta$ of the equation (\ref{Moyal H-P}) 
\be
\label{rozw}
\Theta=\frac{\pi}{2}\cos(p+q)-w\sin q-
\int_{0}^{ z}\sin (\zeta\frac{2}{\hbar}\sin\frac{\hbar}{2}\cdot\cos q+p)d\zeta .
\ee
The $\Theta$ is analytic in $\hbar$, which is the necessary condition for existence of 
$\chi_{\mbox{\tiny{N}}}$. It describes in a classical limit a heavenly metric 
(\ref{example metric}). 

In order to construct a sequence of $su(N)$ chiral fields we assume that the 
coordinates $w,z$ are real. The Cauchy data for (\ref{rozw}) read
\beas
\Theta|_{z=0}&=&\frac{\pi}{2}\cos(p+q)-w\sin q\\
\partial_{z}\Theta|_{z=0}&=&-\sin p
\eeas
As has been described in section \ref{Reduction of ME to $su(N)$ SDYM system} these 
Cauchy data induce the Cauchy data for the $su(N)$ chiral field equation (\ref{chiral})
\beas
\vartheta|_{z=0}&=&\frac{\pi}{2}\,\frac{1}{2}[L_{(1,1)}+(-1)^{N}L_{(N-1,N-1)}]
                   -w\,\frac{1}{2i}[L_{(0,1)}+L_{(0,N-1)}]\\
\partial_{z}\vartheta|_{z=0}&=&-\frac{1}{2i}[L_{(1,0)}+L_{(N-1,0)}]
\eeas
as 
\beas
\chi_{\mbox{\tiny{N}}}(\cos(p+q))&=&\frac{1}{2}[L_{(1,1)}+L_{(-1,-1)}]=\frac{1}{2}[L_{(1,1)}+(-1)^{N}L_{(N-1,N-1)}]\\
\chi_{\mbox{\tiny{N}}}(\sin q)&=&\frac{1}{2i}[L_{(0,1)}-L_{(0,-1)}]=\frac{1}{2i}[L_{(0,1)}+L_{(0,N-1)}]\\
\chi_{\mbox{\tiny{N}}}(\sin p)&=&\frac{1}{2i}[L_{(1,0)}-L_{(-1,0)}]=\frac{1}{2i}[L_{(1,0)}+L_{(N-1,0)}]
\eeas

The expansion of $\Theta$ into Fourier series reads
\bea
\label{Fourier expansion}
\Theta&=&\frac{\pi}{4}[e^{i(p+q)}+e^{-i(p+q)}]-w\frac{1}{2i}[e^{iq}-e^{-iq}]+\\
& &+\sum_{m=1}^{\infty}A_{2m-1}\, \frac{1}{2}[e^{i(p+(2m-1)q)}+e^{-i(p+(2m-1)q)}+e^{i(-p+(2m-1)q)}+e^{-i(-p+(2m-1)q)}]
\nonumber\\
& &+\sum_{m=1}^{\infty}A_{2m}\, \frac{1}{2i}[e^{i(p+2mq)}-e^{-i(p+2mq)}-e^{i(-p+2mq)}+e^{-i(-p+2mq)}]+
A_{0}\, \frac{1}{2i}[e^{ip}-e^{-ip}]\nonumber
\eea
where
\beas
A_{2m-1}&=&\frac{(-1)^{m}}{\frac{2}{\hbar}\sin\frac{\hbar}{2}}\int_{0}^{z\frac{2}{\hbar}\sin\frac{\hbar}{2}}
J_{2m-1}(\zeta)d\zeta\h\h\h m=1,2,...\\
A_{2m}&=&\frac{(-1)^{m+1}}{\frac{2}{\hbar}\sin\frac{\hbar}{2}}\int_{0}^{z\frac{2}{\hbar}\sin\frac{\hbar}{2}}
J_{2m}(\zeta)d\zeta\h\h\h m=0,1,2,...
\eeas
and $J_{\ell}$ is the Bessel function of rank $\ell$.

The formal series of $sl(N,\mbox{\boldmath{$C$}})$ matrices reads
\bea
\label{Fourier after interchange}
\vartheta&=&\frac{\pi}{4}[L_{(1,1)}+L_{(-1,-1)}]-w\frac{1}{2i}[L_{(0,1)}-L_{(0,-1)}]+\\
& &+\sum_{m=1}^{\infty}A_{2m-1}\,\frac{1}{2}[L_{(1,2m-1)}+L_{(-1,-(2m-1))}+L_{(-1,2m-1)}+L_{(1,-(2m-1))}]
\nonumber\\
& &+\sum_{m=1}^{\infty}A_{2m}\,\frac{1}{2i}[L_{(1,2m)}-L_{(-1,-2m)}-L_{(-1,2m)}+L_{(1,-2m)}]+
A_{0}\,\frac{1}{2i}[L_{(1,0)}-L_{(-1,0)}]\nonumber
\eea
The cases of $N$ even or odd are considered separetly.

%
%

\noindent Case 1. $N$ is even. 

From proposition \ref{properties of L} one has

\bi
\item \h $L_{(-1,-1)}=
(-1)^{(N-1+1)(-1)+(N-1+1)(-1)+N(-1)(-1)}L_{(N-1,N-1)}=L_{(N-1,N-1)}$
\item \h $L_{(0,-1)}=\, - L_{(0,N-1)}$
\item \h $L_{(-1,0)}=\, - L_{(N-1,0)}$
\item \h $L_{(1,2m-1)}=\, L_{(1,\mu)+(0,Nr)}=(-1)^{(1+1)r}\,L_{(1,\mu)}=L_{(1,\mu)}$

where $\mu +Nr=2m-1$ and $0\leq\mu\leq N-1$. Thus $\mu=2(m-\frac{N}{2}r)-1$, if one denotes
$\mu=2\nu-1$ then $m=\nu+\frac{N}{2}r$ and the sum 
\mbox{$\sum_{m=1}^{\infty}\mapsto\sum_{\nu=1}^{\frac{N}{2}}\sum_{r=0}^{\infty}$}  
\item \h $L_{(-1,2m-1)}=\,  L_{(N-1,\mu)}$ and $m=\nu+\frac{N}{2}r$ as above
\item \h $L_{(1,-(2m-1))}=\,  L_{(1,N-\mu)}$ 
\item \h $L_{(-1,-(2m-1))}=\,  L_{(N-1,N-\mu)}$ 
\item \h $L_{(1,2m)}=\,  L_{(1,\mu +Nr)}\, =\, L_{(1,\mu)}$ where $\mu=2\sigma$ and 
$\sigma=0,1,...,\frac{N}{2}-1$
\item \h $L_{(-1,2m)}=\, - L_{(N-1,2m)}\, =\, - L_{(N-1,\mu)}$  where $\mu=2\sigma$ and 
\mbox{$\sigma=0,1,...,\frac{N}{2}-1$}
\item \h $L_{(1,-2m)}=\,  L_{(1,-\mu-Nr)}\, =\, L_{(1,N-\mu)}$  where $\mu=2\sigma$ and 
$\sigma=0,1,...,\frac{N}{2}-1$
\item \h $L_{(-1,-2m)}=\, - L_{(N-1,N-\mu)}$  where $\mu=2\sigma$ and 
$\sigma=0,1,...,\frac{N}{2}-1$
\ei
Substituting all that into (\ref{Fourier after interchange}) one gets
\bea
\label{N even}
\vartheta&=&\frac{\pi}{2}\underbrace{\frac{1}{2}(L_{(1,1)}+L_{(N-1,N-1)})}-
w\underbrace{\frac{1}{2i}(L_{(0,1)}+L_{(0,N-1)})}+\\
& &+\sum_{\nu=1}^{\frac{N}{2}}a_{2\nu-1}[\underbrace{\frac{1}{2}(
L_{(1,2\nu-1)}  +   L_{(N-1,N-(2\nu-1))})}  +   
\underbrace{\frac{1}{2}(L_{(N-1,2\nu-1)}  +  L_{1,N-(2\nu-1)})}]\nonumber\\
& &+\sum_{\nu=1}^{\frac{N}{2}-1}a_{2\nu}[\underbrace{\frac{1}{2i}(
L_{(1,2\nu)}+L_{(N-1,N-2\nu)})}+
\underbrace{\frac{1}{2i}(L_{(N-1,2\nu)}+L_{(1,N-2\nu)})}]\nonumber\\
& &+a_{0}\underbrace{\frac{1}{2i}(L_{(1,0)}+L_{(N-1,0)})}\nonumber
\eea
where
\beas
a_{2\nu-1}&=&
\frac{(-1)^{\nu}}{\frac{N}{\pi}\sin\frac{\pi}{N}}
\sum_{r=0}^{\infty}(-1)^{\frac{N}{2}r}
\int_{0}^{z\frac{N}{\pi}\sin\frac{\pi}{N}}J_{2\nu-1+Nr}(\zeta)d\zeta,
\h\h\h\nu=1,...,\frac{N}{2}\\
a_{2\nu}&=&
\frac{(-1)^{\nu+1}}{\frac{N}{\pi}\sin\frac{\pi}{N}}
\sum_{r=0}^{\infty}(-1)^{\frac{N}{2}r}
\int_{0}^{z\frac{N}{\pi}\sin\frac{\pi}{N}}J_{2\nu+Nr}(\zeta)d\zeta,
\h\h\h\nu=1,...,\frac{N}{2}-1\\
a_{0}&=&
\frac{-1}{\frac{N}{\pi}\sin\frac{\pi}{N}}\,(\,2
\sum_{r=0}^{\infty}(-1)^{\frac{N}{2}r}
\int_{0}^{z\frac{N}{\pi}\sin\frac{\pi}{N}}J_{Nr}(\zeta)d\zeta+
\int_{0}^{z\frac{N}{\pi}\sin\frac{\pi}{N}}J_{0}(\zeta)d\zeta\,)
\eeas
%
%

\vspace{2 ex}

\noindent Case 2. $N$ is odd. 

In this case 

\bi
\item \h $L_{(-1,-1)}=
(-1)^{(N-1+1)(-1)+(N-1+1)(-1)+N(-1)(-1)}L_{(N-1,N-1)}=-L_{(N-1,N-1)}$
\item \h $L_{(0,-1)}=\, - L_{(0,N-1)}$
\item \h $L_{(-1,0)}=\, - L_{(N-1,0)}$
\item \h $L_{(1,2m-1)}=\, L_{(1,\mu)+(0,Nr)}=(-1)^{(1+1)r}\,L_{(1,\mu)}=L_{(1,\mu)}$

where $\mu +Nr=2m-1$ and $0\leq\mu\leq N-1$.
\ei 
Observe that if $\mu$ is odd then $r$ must be even and vice versa if $\mu$ is even
then $r$ must be odd.  We divide the sum 
\mbox{$\sum_{m=1}^{\infty}\mapsto\sum_{\mu=1}^{N-1}\sum_{r=0}^{\infty}$}
into  two separete sums
\beas
\left\{\begin{array}{ll}
\mu=2\nu-1 & \nu=1,...,\frac{N-1}{2}\\
r=2k & k=0,1,...,\infty\\
m=\nu+Nk  & \end{array}\right.\h\h
\left\{\begin{array}{ll}
\mu=2\nu & \nu=0,1,...,\frac{N-1}{2}\\
r=2k+1 & k=0,1,...,\infty\\
m=\nu+\frac{N+1}{2}+Nk  & \end{array}\right.
\eeas

\bi
\item \h $L_{(-1,2m-1)}=\, (-1)^{r} L_{(N-1,\mu)}$ and $m=\nu+\frac{N}{2}r$ as above
\item \h $L_{(1,-(2m-1))}=\,  L_{(1,N-\mu)}$ 
\item \h $L_{(-1,-(2m-1))}=\,(-1)^{r+1}  L_{(N-1,N-\mu)}$ 
\item \h $L_{(1,2m)}=\,  L_{(1,\mu +Nr)}\, =\, L_{(1,\mu)}$ 
\ei
where this time $2m=\mu+Nr$ and $\mu$ odd implies $r$ odd and $\mu$ even implies $r$ even.  
\beas
\left\{\begin{array}{ll}
\mu=2\sigma-1 & \sigma=1,...,\frac{N-1}{2}\\
r=2l+1 & l=0,1,...,\infty\\
m=\sigma + \frac{N-1}{2}+Nl  & \end{array}\right.\h\h
\left\{\begin{array}{ll}
\mu=2\sigma & \sigma=1,...,\frac{N-1}{2}\\
r=2l & l=0,1,...,\infty\h\h\mbox{if} \\
(\h \sigma=0 & \Rightarrow\h l\neq 0\h )\\
m=\sigma+Nl  & \end{array}\right.
\eeas

\bi
\item \h $L_{(-1,2m)}=\, - L_{(N-1,2m)}\, =\, (-1)^{r+1} L_{(N-1,\mu)}$  
\item \h $L_{(1,-2m)}=\,  L_{(1,-\mu-Nr)}\, =\, L_{(1,N-\mu)}$  
\item \h $L_{(-1,-2m)}=\, (-1)^{r} L_{(N-1,N-\mu)}$  
\ei
Substituting all that into (\ref{Fourier after interchange}) we obtain
\bea
\label{N odd}
\vartheta&=&\frac{\pi}{2}\underbrace{\frac{1}{2}(L_{(1,1)}-L_{(N-1,N-1)})}-w
\underbrace{\frac{1}{2i}(L_{(0,1)}+L_{(0,N-1)})}+\\
& &+\sum_{\nu=1}^{\frac{N-1}{2}}a_{2\nu-1}\,
[\underbrace{\frac{1}{2}(L_{(1,2\nu-1)}-L_{(N-1,N-(2\nu-1))})} + 
\underbrace{\frac{1}{2}(L_{(N-1,2\nu-1)}+L_{(1,N-(2\nu-1))})}]\nonumber\\
& &+\sum_{\nu=1}^{\frac{N-1}{2}}a_{2\nu}
[\underbrace{\frac{1}{2}(L_{(1,2\nu)}+ L_{(N-1,N-2\nu)})} + 
\underbrace{\frac{1}{2}(L_{(1,N-2\nu)}-L_{(N-1,2\nu)})}]\nonumber\\
& &+\sum_{\nu=1}^{\frac{N-1}{2}}b_{2\nu-1}
[\underbrace{\frac{1}{2i}(L_{(1,2\nu-1)}+L_{(N-1,N-(2\nu-1))})} + 
\underbrace{\frac{1}{2i}(L_{(1,N-(2\nu-1))} - L_{(N-1,2\nu-1)})}]\nonumber\\
& &+\sum_{\nu=1}^{\frac{N-1}{2}}b_{2\nu}
[\underbrace{\frac{1}{2i}(L_{(1,2\nu)} - L_{(N-1,N-2\nu)})} + 
\underbrace{\frac{1}{2i}(L_{(N-1,2\nu)}+L_{(1,N-2\nu)})}]\nonumber\\
& &+2a_{0}\underbrace{\frac{1}{2}[L_{(1,0)}-L_{(N-1,0)}]}
+b_{0}\underbrace{\frac{1}{2i}[L_{(1,0)}+L_{(N-1,0)}]}\nonumber
\eea
where
\beas
a_{2\nu-1}&=&
\frac{(-1)^{\nu}}{\frac{N}{\pi}\sin\frac{\pi}{N}}
\sum_{k=0}^{\infty}(-1)^{k} 
\int_{0}^{z\frac{N}{\pi}\sin\frac{\pi}{N}}J_{2\nu-1+N2k}(\zeta)d\zeta\h\h\h\nu=1,2,...,\frac{N-1}{2}\\
a_{2\nu}&=&
\frac{(-1)^{\nu+\frac{N+1}{2}}}{\frac{N}{\pi}\sin\frac{\pi}{N}}
\sum_{k=0}^{\infty}(-1)^{k}
\int_{0}^{z\frac{N}{\pi}\sin\frac{\pi}{N}}J_{2\nu+N(2k+1)}(\zeta)d\zeta\h\h\h\nu=0,1,2,...,\frac{N-1}{2}\\
b_{2\nu-1}&=&
\frac{(-1)^{\nu+\frac{N+1}{2}}}{\frac{N}{\pi}\sin\frac{\pi}{N}}
\sum_{k=0}^{\infty}(-1)^{k}
\int_{0}^{z\frac{N}{\pi}\sin\frac{\pi}{N}}J_{2\nu-1+N(2k+1)}(\zeta)d\zeta\h\h\h\nu=1,2,...,\frac{N-1}{2}\\
b_{2\nu}&=&
\frac{(-1)^{\nu+1}}{\frac{N}{\pi}\sin\frac{\pi}{N}}
\sum_{k=0}^{\infty}(-1)^{k}
\int_{0}^{z\frac{N}{\pi}\sin\frac{\pi}{N}}J_{2\nu+N2k}(\zeta)d\zeta\h\h\h\nu=1,2,...,\frac{N-1}{2}\\
b_{0}&=&-
\frac{1}{\frac{N}{\pi}\sin\frac{\pi}{N}}\,[2
\sum_{k=1}^{\infty}(-1)^{k}
\int_{0}^{z\frac{N}{\pi}\sin\frac{\pi}{N}}J_{2Nk}(\zeta)d\zeta +
\int_{0}^{z\frac{N}{\pi}\sin\frac{\pi}{N}}J_{0}(\zeta)d\zeta]  
\eeas
In the formulas (\ref{N even}) and (\ref{N odd}) the linear combinations denoted by underbraces
are elements of the basis of $su(N)$ for $N$ even or odd, respectively.

Finally (\ref{N even}) and (\ref{N odd}) give the sequence of $su(N)$ 
chiral fields tending for $N\rightarrow\infty$ to the heavenly space described by 
the metric (\ref{example metric}).

In particular, substituting $N=2$ into (\ref{N even}) and applying the formulas
\beas
\sum_{k=1}^{\infty}(-1)^{k}
J_{2k+1}(\zeta)&=&\frac{\sin\zeta}{\zeta}\\
2\sum_{k=1}^{\infty}(-1)^{k}
J_{2k}(\zeta)+
J_{0}(\zeta)&=&\cos\zeta
\eeas
and the fact that
\[
L_{(1,1)}=-\frac{i}{\pi}\sigma_{1}\h ,\h\h
\frac{1}{i}L_{(0,1)}=\frac{i}{\pi}\sigma_{2}\h ,\h\h
\frac{1}{i}L_{(1,0)}=\frac{i}{\pi}\sigma_{3}
\]
where $\sigma_{1},\sigma_{2},\sigma_{3}$ are the Pauli matrices one gets 
\be
\vartheta=\frac{1}{2i}\cos(\frac{2}{\pi}z)\sigma_{1}+
\frac{w}{\pi i}\sigma_{2}+
\frac{1}{2i}\sin(\frac{2}{\pi}z)\sigma_{3}.
\ee
\eexam
\section*{Conclusions}
The purpose of this paper was to justify the name \it master equation \rm (ME) for 
equation (\ref{ME in H}). In particular we were able to show (in theorem \ref{o homomorfizmie
rozwiazan}) that any analytic solution of the $sl(N,\mbox{\boldmath{$C$}})$ equation 
could be obtained from a suitable solution of ME. This embedding holds true also for the 
solutions of the HP heavenly equation as well as for the solutions of the 
$sl(N,\mbox{\boldmath{$C$}})$ chiral equation in 2 diemensions or Ward's integrable 
chiral equations in $2+1$ diemensions (Ward 1988, Ward 1995, Przanowski and Forma\'{n}ski
1999, Dunajski and Manton 2005).

Moreover, as has been shown in section \ref{From chiral fields to heavenly space} 
there exists a natural method to construct sequences of su(N) (or $sl(N,\mbox{\boldmath{$C$}})$)
chiral fields tending to heavenly spaces when $N\rightarrow\infty$. It would be very 
interesting  to apply the  described procedure to the finite energy solutions of chiral field
equation (\it unitons \rm ; Uhlenbeck (1989), Ward (1990a)) or Ward's integrable chiral equations 
in $2+1$ diemensions. As a result one would obtain a heavenly space. The question 
arises wheather this space is one of a finite action (an \it instanton\rm ). 

Another interesting open problem is to try to embed the \it Kadomtsev-Petviashvili \rm (KP) 
equation into ME. This posibility is suggested by (Mason 1990, Mason and Woodhouse 1996, Strachan 1997,
Przanowski and Forma\'{n}ski 1999). We expect that the extension of the Poisson algebra to 
its quantum deformation is sufficient to encode KP equation in ME.

Of cource the main technical problem is to find an effective method of looking for solutions
of ME.
\section*{Acknowledgements}
We are indebted to Maciej Dunajski and Jacek Tafel for the interest in this work and a very 
fruitful comments.
The work was partially supported by CONACyT grant no: 41993-F (Mexico) and NATO grant no: 
PST.CLG.978984

\end{document}